\documentclass[aps,prx,twocolumn,noshowpacs,superscriptaddress,preprintnumbers,amsmath,amssymb]{revtex4-1}

\usepackage{graphicx}
\usepackage{epsfig}
\usepackage{dcolumn}
\usepackage{bm}
\usepackage{longtable}
\usepackage{color}
\usepackage{bm}
\usepackage{epstopdf}

\begin{document}


\title{Role of spin-orbit coupling and evolution of the electronic structure of WTe$_2$ under an external magnetic field}



\author{D.\ Rhodes}
\affiliation{National High Magnetic Field Laboratory, Florida
State University, Tallahassee-FL 32310, USA}
\author{S.\ Das}
\affiliation{Department of Physics and National High Magnetic Field Laboratory, Florida
State University, Tallahassee-FL 32310, USA}
\author{Q.\ R.\ Zhang}
\affiliation{National High Magnetic Field Laboratory, Florida
State University, Tallahassee-FL 32310, USA}
\author{B.\ Zeng}
\affiliation{National High Magnetic Field Laboratory, Florida
State University, Tallahassee-FL 32310, USA}
\author{N.\ R.\ Pradhan}
\affiliation{National High Magnetic Field Laboratory, Florida
State University, Tallahassee-FL 32310, USA}
\author{N.\ Kikugawa}
\affiliation{National Institute for Materials Science, Tsukuba, Ibaraki 305-0047, Japan}
\affiliation{National High Magnetic Field Laboratory, Florida
State University, Tallahassee-FL 32310, USA}
\author{E.\ Manousakis}
\affiliation{Department of Physics and National High Magnetic Field Laboratory, Florida
State University, Tallahassee-FL 32310, USA}
\affiliation{Department of Physics, University of Athens, Panepistimioupolis, Zografos, 157 84 Athens, Greece}
\author{L.\ Balicas}\email{balicas@magnet.fsu.edu}
\affiliation{National High Magnetic Field Laboratory, Florida
State University, Tallahassee-FL 32310, USA}


\date{\today}

\begin{abstract}
Here, we present a detailed study on the temperature and angular dependence of the Shubnikov-de-Haas (SdH) effect in the semi-metal WTe$_2$. This compound was recently shown to display a very large non-saturating magnetoresistance which was attributed to nearly perfectly compensated densities of electrons and holes. We observe four fundamental SdH frequencies and attribute them to spin-orbit split, electron- and hole-like, Fermi surface (FS) cross-sectional areas. Their angular dependence is consistent with ellipsoidal FSs with volumes implying a modest excess in the density of electrons with respect to that of the holes. We show that density functional theory (DFT) calculations fail to correctly describe the FSs of WTe$_2$. When their cross-sectional areas are adjusted to reflect the experimental data, the resulting volumes of the electron/hole FSs obtained from the DFT calculations would imply a pronounced imbalance between the densities of electrons and holes. We find evidence for field-dependent Fermi surface cross-sectional areas by fitting the oscillatory component superimposed onto the magnetoresistivity signal to several Lifshitz-Kosevich components. We also observe a pronounced field-induced renormalization of the effective masses. Taken together, our observations suggest that the electronic structure of WTe$_2$ evolves with the magnetic field. This evolution might be a factor contributing to its pronounced magnetoresistivity.
 \end{abstract}


\keywords{Subject Areas: Condensed Matter Physics}
\maketitle

\section{Introduction}
Transition metal dichalcogenides \emph{TM}X$_2$ (where \emph{TM} stands for a transition metal, and X for a chalcogen) have become the subject of intense research activity since, as graphene, they can be exfoliated down to a single atomic layer \cite{reviews}. Semiconducting compounds present remarkable optoelectronic properties\cite{kis} while the whole family displays a wide range of electronic states \cite{reviews}; from wide gap semiconductors such as trigonal 1\emph{T}-HfS$_2$\cite{gaiser}, to band gaps comparable to Si such as in $\alpha$-MoTe$_2$\cite{heinz}, or Mott systems  such as 1\emph{T}-TaS$_2$\cite{forro}. Some compounds display charge-density wave(s) that coexist with superconductivity as in \emph{H}-NbSe$_2$\cite{NbSe2}, while others are semi-metals such as $\beta$-MoTe$_2$ or WTe$_2$\cite{cava}.

WTe$_2$, which crystallizes in a distorted, orthorhombic variant of the CdI$_2$ structure with octahedral coordination around W, was recently reported to display an extremely pronounced and non-saturating magnetoresistance which\cite{cava}. Following simple arguments already discussed by Pippard\cite{pippard}, this effect was attributed to a nearly perfect compensation between the density of electrons and holes. Evidence for electron/hole compensation was recently provided by angle-resolved photoemission (ARPES) experiments claiming to observe electron- and hole-like Fermi surface pockets of almost the same size\cite{pletikosic}. These results contrast markedly with a subsequent ARPES study\cite{jiang} which claim to observe up to 9 FS sheets, including a hole-pocket at the Brillouin zone center or the $\Gamma$-point, and two hole-pockets and two electron-pockets on each side of $\Gamma$ and along the $\Gamma$-\emph{X} direction. The same study finds evidence for spin-orbit split bands and for an opposite texture between the spin and the angular orbital momentum around the $\Gamma$-point which should suppress carrier backscattering\cite{jiang}. The application of a magnetic field would not only alter the spin texture, it would also displace, through the Zeeman effect, the spin-split bands relative to the Fermi level, opening backscattering channels and hence increasing the resistivity. Similar arguments were invoked to explain the pronounced magnetoresistivity observed in Cd$_3$As$_2$\cite{ong}.

Electrical transport studies focused on the Shubnikov-de Haas-effect yielding results \cite{xiang_three_pockets, behnia, zhao, cai} which contrast markedly with both ARPES studies. The first study reports the observation of just 3 Fermi surface pockets in marked contrast with band structure calculations which always predict an even number of FS pockets for bulk WTe$_2$\cite{pletikosic,behnia,Lv}. Subsequent studies, more in line with the calculations, observe 4 Fermi surface pockets, or two sets of concentric electron- and hole-like FSs\cite{behnia,zhao,cai} at either side of the $\Gamma-$point. The angular dependence of the observed frequencies, or of the FS cross-sectional areas obtained from the Onsager relation, suggests ellipsoidal pockets with volumes yielding an excess of 4 \% in the density of holes when compared that of the electrons. This imbalance would be comparable to that of elemental Bi which, in contrast to WTe$_2$, displays a saturating magnetoresistance\cite{fauque}.

Subsequent calculations \cite{bernevig} confirm the previously reported FS geometry \cite{pletikosic,behnia,Lv}, while analyzing the topological properties of its electronic structure around the Fermi level. Ref. \onlinecite{bernevig} finds a linear touching between hole and electron Fermi-surfaces leading to Berry phase curvature singularities, or to two sets of four Weyl-points in the first Brillouin zone, each Weyl-point characterized by topological charges of $\pm 1$. At the moment it remains unclear how such electronic structure would contribute to WTe$_2$ pronounced magnetoresistivity. Ref. \onlinecite{bernevig} also finds that its FS is particularly sensitive to the exact position of the Fermi level $\varepsilon_F$, with small displacements in $\varepsilon_F$ leading to topological Lifshitz transitions, in agreement with our DFT calculations being discussed within this manuscript.

Here, we present angle- and temperature-dependent magnetoresistivity measurements in fields up to $H=35$ T on WTe$_2$ single-crystals, in an attempt to elucidate the number of Fermi surface sheets. In parallel, we performed density-functional theory (DFT) calculations incorporating the spin-orbit effect. By adjusting the position of the chemical potential we attempted to precisely describe the geometry of the measured Fermi surfaces. In agreement with a previous report\cite{behnia} we observe four fundamental Shubnikov-de-Haas frequencies, or Fermi surface cross-sectional areas, which we ascribe to spin-orbit split electron- and hole-pockets. As argued in Ref.~\onlinecite{behnia}, their angular dependence
can be described in terms of ellipsoidal FSs. Therefore, we find that the DFT calculations are unable to correctly describe the angular dependence of the FS cross-sectional areas, particularly in what concerns the hole-pockets. For fields aligned along the $c-$ and the $b-$axis we observe the effective masses to become renormalized, i.e. considerably heavier as the field increases. As the FFT spectrum of the oscillatory signal, taken on a limited field window, is unable to resolve all cross-sectional areas, we
fit it to five and four Lifshitz-Kosevich components. Our fits clearly indicate that one cannot fit the SdH signal over the entire field range by maintaining the frequencies and the effective masses as constants. Good fits are obtained within a limited field range, confirming that the effective masses (and likely the cross-sectional areas) are field-dependent.

\section{Experimental}
WTe$_2$ was grown via a chemical vapor transport technique using chlorine as the transport agent: W (99.999 \%) and Te (99.9999 \%) were heated up to a peak temperature of 750 $^{\circ}$C at a rate of 100 $^{\circ}$C/h, and subsequently cooled down to room temperature at a rate of 10 $^{\circ}$C/h. The obtained powder was ground with a mortar and pestle and heated again following the the same temperature profile. The remaining product was subsequently combined with TeCl$_4$ (99.999 \%, 1.5 mg/cm$^3$) and heated up to a peak temperature of 750 $^{\circ}$C at a rate of 100 $^{\circ}$C/h under a temperature gradient of $\approx$ 50 $^{\circ}$C. The crystals used for this study displayed resistivity ratios in excess of 10$^{2}$. Subsequent crystals grown via the Te flux method, yielded much higher RRRs ($\sim 700$) and show similar behavior (see supplementary information).

The synthesis procedure yielded platelet like single crystals, several millimeters in length and typically a millimeter wide, making the \emph{a-}, \emph{b-}, and the \emph{c}-axis easily distinguishable. The stoichiometric composition was determined by energy dispersive \emph{X}-ray spectroscopy and single-crystal \emph{X}-ray refinement. A conventional four terminal configuration was used for the resistivity measurements which were performed under high magnetic-fields either by using a 35 T resistive magnet or a Physical Property Measurement System (PPMS).

\begin{figure}[htb]
\begin{center}
\includegraphics[width = 8.2 cm]{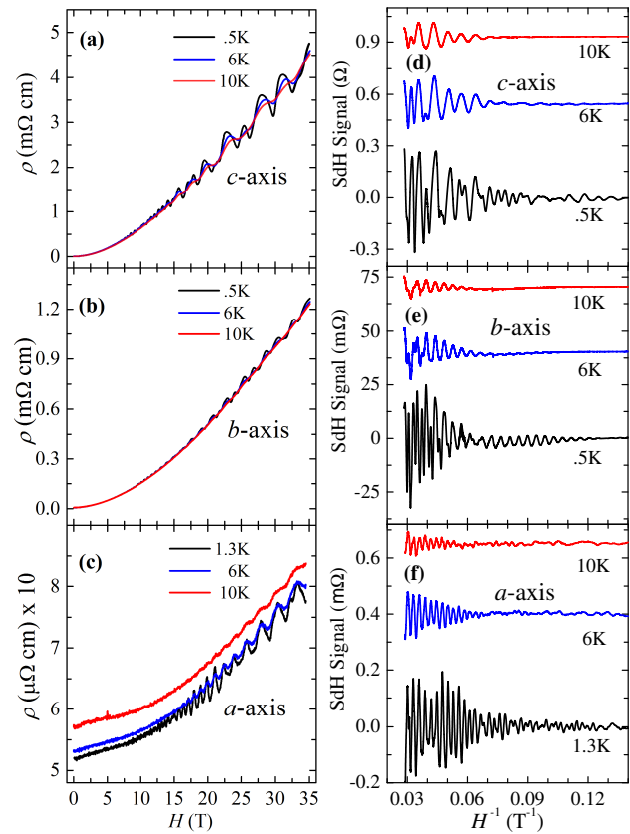}
\caption{(a)Resistivity $\rho$ as a function of the magnetic field $H$ applied along the \emph{c}-axis, for a WTe$_2$ single-crystal and for three representative values of the temperature $T$, i.e. 0.5 K (black line), 6 K (blue) and 10 K (red), respectively. Here, the current is applied along the \emph{a}-axis, hence perpendicularly to $H$. (b)Same as in (a) but for fields along the \emph{b}-axis of the crystal. (c) Same as in (a) but for currents parallel to $H$ which is applied along the \emph{a}-axis. (d) Shubnikov-de-Haas signal superimposed onto the curves in (a) after subtraction of a polynomial background. (e) Same as in (d) but from the curves in (b). (f) Same as in (d) but from the curves in (c).}
\label{fig:1}
\end{center}
\end{figure}
Figures \ref{fig:1} (a), (b), and (c) show the resistivity $\rho$ of a WTe$_2$ single-crystal as a function of the magnetic field applied along the \emph{a}, \emph{b}, and \emph{c}-axis respectively, for several temperatures and with the electrical current applied along the \emph{a}-axis.  In agreement with previous reports\cite{cava,behnia}, fields along the \emph{c}-axis yield the largest magnetoresistivity, although in our crystals the anisotropy in magnetoresistivity between fields along the \emph{c-} and the \emph{b-}axis is just a factor of 4 and not beyond one order of magnitude\cite{behnia}. The smallest change in resistivity is observed for fields applied along the \emph{a}-axis, since in this configuration the Lorentz force is expected to vanish. For all three sets of data the oscillatory component superimposed onto the smoothly increasing background corresponds to the Shubnikov-de-Haas effect. Figures \ref{fig:1} (d), (e), (f), display  the oscillatory component superimposed onto $\rho(H,T)$ as a function of the inverse field $H^{-1}$, for fields oriented along all three crystallographic axes and for several temperatures (from the curves in (a), (b) and (c), respectively). To obtain the oscillatory component, the smoothly varying background was fit to a polynomial which was subsequently subtracted from the raw data.

\begin{figure}[htb]
\begin{center}
\includegraphics[width = 7.5 cm]{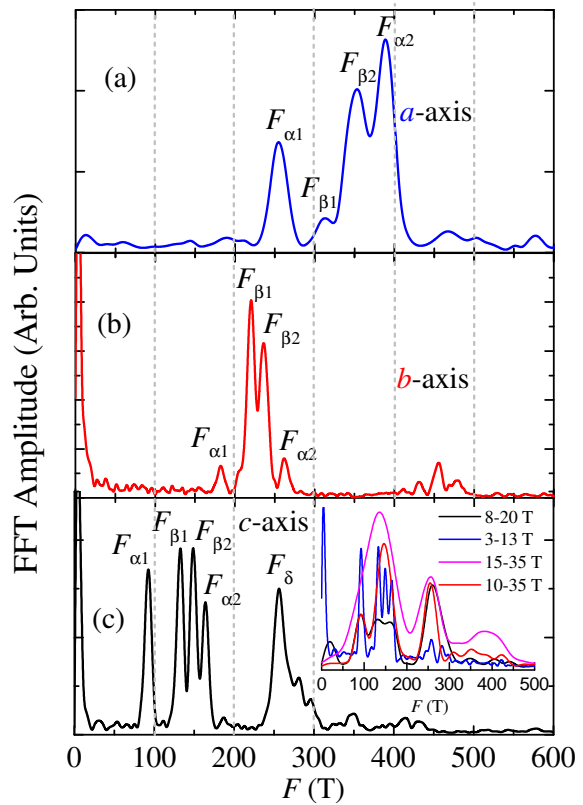}
\caption{(a) Fast Fourier transform of the SdH signal superimposed onto $\rho(H)$ for fields along the \emph{a}-axis and for $T = 1.3$ K. Here, the oscillatory signal was extracted from a field interval ranging from 3 to 35 T. As is discussed in the main text, the indexation of the four peaks observed in the FFT spectrum is based upon band structure calculations, where $F_{\alpha}$ and $F_{\beta}$ denotes hole- and electron-pockets respectively, and the indexes 1 and 2 differentiate spin-orbit split Fermi surface pockets. (b) Same as in (a) but for fields along the \emph{b}-axis and for $T=0.5$ K. (c) Same as in (a) but for fields along the \emph{c}-axis and for $T=0.5$ K. Inset: FFT spectra extracted for different field intervals. Notice the loss of frequency resolution when the FFT is extracted either from traces acquired under high fields or from a limited field range.
To allow the visual comparison between the different FFT spectra the blue and black traces were multiplied by a factor 75 and 3.5, respectively.
Notice how within the interval $\Delta H = 8-20 $ T the peak corresponding to the magnetic breakdown orbit $F_{\delta}$ becomes more pronounced than the fundamental peaks.}
\label{fig:2}
\end{center}
\end{figure}
\begin{figure*}[htb]
\begin{center}
\includegraphics[width = 12 cm]{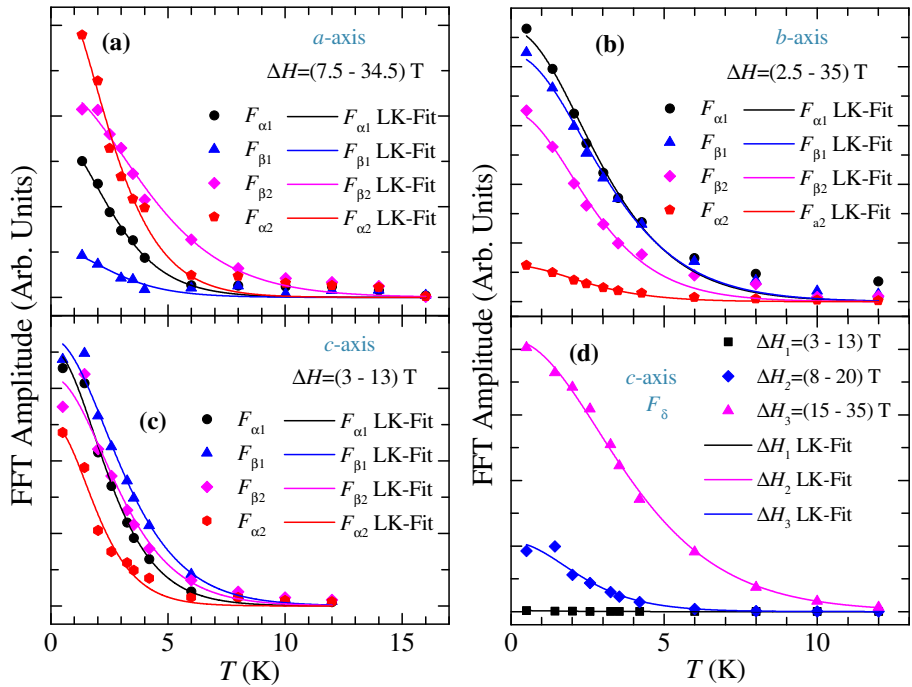}
\caption{(a) Magnitude of the FFT peaks extracted from the oscillatory component superimposed into the magnetoresistivity measured for fields applied along the \emph{a}-axis and as a function of the temperature \emph{T}. Solid lines are linear fits to the Lifshitz-Kosevich formalism from which we extract the effective masses $\mu$. (b) Same as in (a) but for fields along the \emph{b}-axis. (c) Same as in (a) but for fields along the \emph{c}-axis. (d) Magnitude of FFT peak observed at $F_{\delta}$ which corresponds to a magnetic breakdown orbit.}
\label{fig:3}
\end{center}
\end{figure*}

Figures \ref{fig:2} (a), (b), and (c) show the fast Fourier transform (FFT) of the SdH oscillatory signal displayed in Figs.~\ref{fig:1} (d), (e), and (f), respectively.
As seen, for fields along all three crystallographic axes four fundamental frequencies are clearly observed. This is in agreement with the observations of
Ref.~\onlinecite{behnia} but contrasts markedly with both ARPES studies \onlinecite{pletikosic,jiang}. For fields along the $c$-axis one observes peak frequencies at  \emph{F}$_{\alpha1} = 90$ T, \emph{F}$_{\beta1} = 130$ T, \emph{F}$_{\beta2} = 145$ T, and \emph{F}$_{\alpha2} = 160$ T, respectively. As we discuss below, and based on DFT calculations, the $\alpha$ and the $\beta$ frequencies are identified with spin-orbit split hole- and electron-pockets, respectively. The indexes 1 and 2 distinguish the smaller from the larger spin-orbit split Fermi surfaces. The DFT calculations indicate the presence of identical sets of electron and hole pockets at either side of the $\Gamma-$point making it a total of 8 Fermi surface sheets in the first Brillouin zone. We see a gradual increase in the frequencies as the crystal is rotated from \emph{c}- towards the \emph{b}-axis or the \emph{a}-axis. For fields along the \emph{b}-axis the peaks are shifted towards \emph{F}$_{\alpha1} = 180$ T, \emph{F}$_{\beta1} = 220$ T, \emph{F}$_{\beta2} = 236$ T, and \emph{F}$_{\alpha2} =262$ T, respectively. For fields along the \emph{a}-axis the peaks are observed at \emph{F}$_{\alpha1} = 250$ T, \emph{F}$_{\beta1} = 300$ T, \emph{F}$_{\beta2} = 357$ T, and \emph{F}$_{\alpha2}= 390$ T, respectively. Ref. \onlinecite{jiang} claims the existence of 9 Fermi surface pockets in contrast to the eight pockets implied by our experiments when combined with the calculations described below. Here, we just mention that another very small electron pocket do emerge at the $\Gamma-$point by shifting the Fermi level in the DFT calculations with respect to the position required to match our observations. Indeed this would lead to a total number of 9 Fermi surface sheets. Nevertheless, their cross-sectional areas would strongly disagree with our experimental observations. Furthermore, the different ARPES studies \cite{pletikosic,jiang} disagree among themselves on the existence of this extra pocket. As we show below, the DFT calculations indicate that the geometry of the Fermi surface(s) is extremely sensitive with respect to small displacements of the Fermi level, i.e. of the order of 20 meV, which is comparable to the experimental resolution of ARPES. The distinction among the ARPES results is attributable to the resolution between the different set-ups.

For fields along the \emph{c}-axis one also observes a pronounced fifth peak, \emph{F}$_{\delta} \simeq 250$ T, which clearly corresponds to $F_{\alpha1}+ F_{\alpha2}$. This frequency was previously attributed to magnetic breakdown or to the magnetic field-induced carrier tunneling between both orbits\cite{behnia}. However, the fact that the amplitude of the peak associated with \emph{F}$_{\delta}$ is more pronounced than peaks associated with the fundamental orbits, such as $F_{\alpha2}$, contradicts this scenario, unless both FS sheets were
 in close proximity at the point of nearly touching. In fact, as seen in the inset of Fig.~\ref{fig:2} (c), which displays the FFT spectra acquired under distinct field intervals, this orbit becomes progressively more pronounced at higher fields, which is consistent with magnetic breakdown. However, as illustrated by the FFT spectrum extracted from the field interval ranging from 8 to 20 T, the amplitude of $F_{\delta}$ can, in some field intervals, become more pronounced than the amplitude of all other fundamental peaks. This suggests that the Zeeman-effect is altering the geometry of the $\alpha$-pockets, hence decreasing their (spin-orbit induced) splitting in some region(s) of \emph{k}-space. Our DFT analysis, discussed below, is indeed consistent with this possibility. This would imply that the electronic structure at the Fermi level evolves with field due to the Zeeman-effect. We considered this hypothesis and evaluated the possibility of studying in detail the evolution of the SdH frequencies, and of the concomitant effective masses, as a function of the field. But as illustrated by the inset of Fig.~\ref{fig:2} (c), the resolution among peaks in the FFT spectra, or the peak sharpness, decreases considerably when the very low field interval is excluded from the oscillatory signal, thus preventing such a detailed analysis.
\begin{figure*}[htb]
\begin{center}
\includegraphics[width = 16 cm]{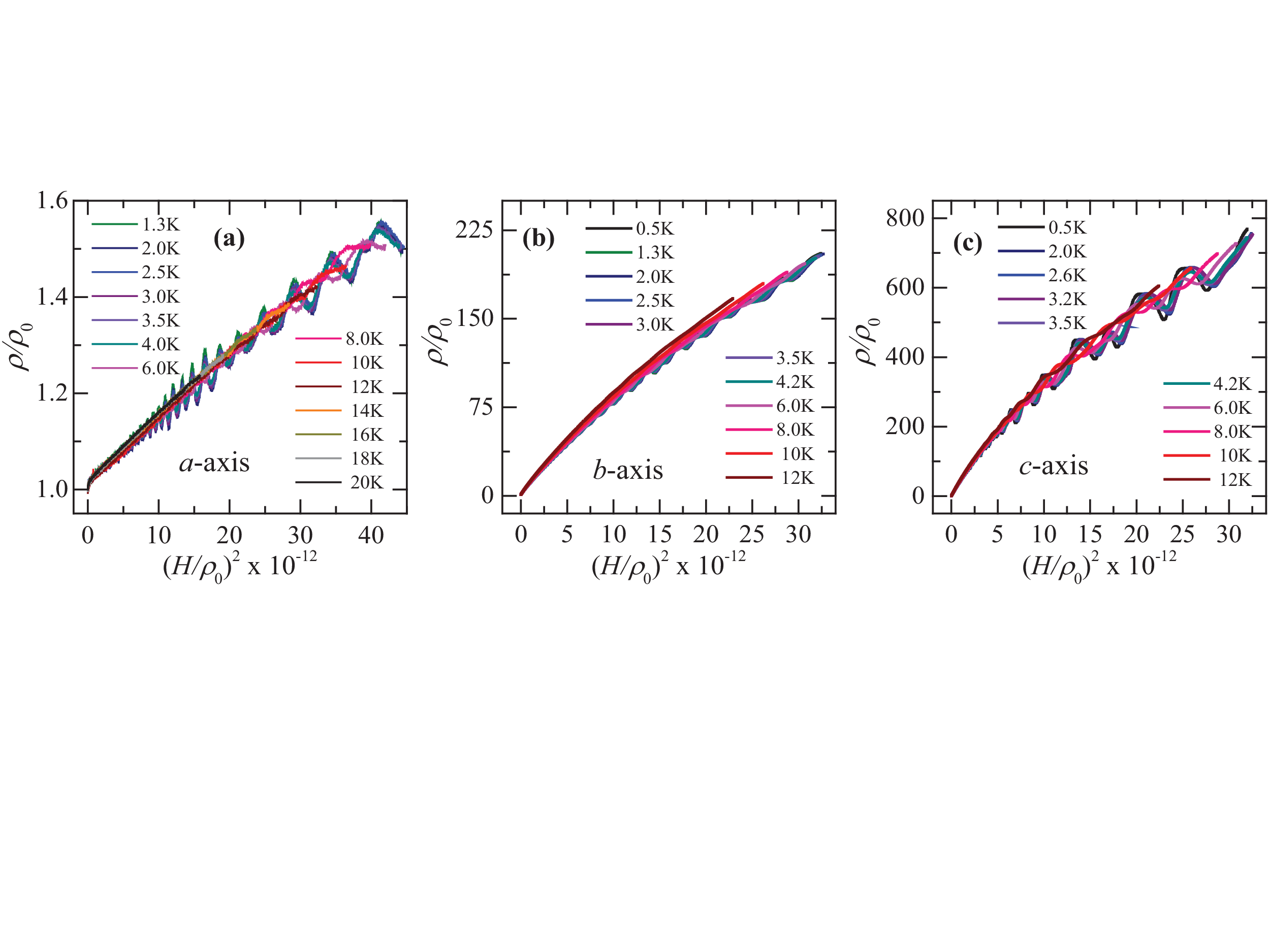}
\caption{(a) Kohler plot of the resistivity $\rho$, normalized by its zero field value $\rho_\text{0}$, as a function of $(H/\rho_\text{0})^2$ for fields applied along the $a-$axis. (b) Same as in (a) but for fields along the \emph{b-}axis. (c) Same as in (a) but for fields along the \emph{c-}axis.}
\label{fig:4}
\end{center}
\end{figure*}

Figures \ref{fig:3} (a), (b), and (c) display the amplitude of each peak observed in the FFT spectra as a function of the temperature $T$, for fields along the $a-$, $b-$, and the $c-$axis, respectively. Solid lines are fits to the Lifshitz-Kosevich formula $A/\sinh X$ (where $X = 2 \pi^2k_B T/\hbar\omega_c = 14.69 \mu T/\overline{H}$, with $\mu$ being the effective mass $\mu$ and $\overline{H}$ the average field value), from which we extract the effective masses. As seen, the fits are excellent and yield precise values for $\mu$. More importantly, according to the fits in Fig.~\ref{fig:3} (d), the effective mass of $F_{\delta}$, or $\mu_{\delta}$, is found to evolve as a function of the field: i.e. $\mu_{\delta} = (0.73 \pm 0.05)m_0$ for $H$ ranging from 3 to 13 T, $\mu_{\delta} =(0.9 \pm 0.07)m_0$ for fields between 8 and 20 T, $\mu_{\delta} =(0.97 \pm 0.02)m_0$ for fields ranging from 15 to 35 T. In Table I below summarizes the experimental information extracted from this study, including effective masses for $H\parallel c-$ and to the $b-$axis, in each case extracted from two intervals in field, i.e. a low field region and the whole field window up to 35 T. As seen, the effective masses extracted from the whole field range are approximately a factor of 2 heavier than those extracted from the low field region, clearly indicating a renormalization of the  effective mass induced by the application of a magnetic field. In the same table we include the results of an attempt to match the experimentally determined FSs with those determined from the DFT calculations by independently adjusting the position of the electron- and hole-bands relative to the Fermi level.
\begin{figure*}[htb]
\begin{center}
\includegraphics[width = 13 cm]{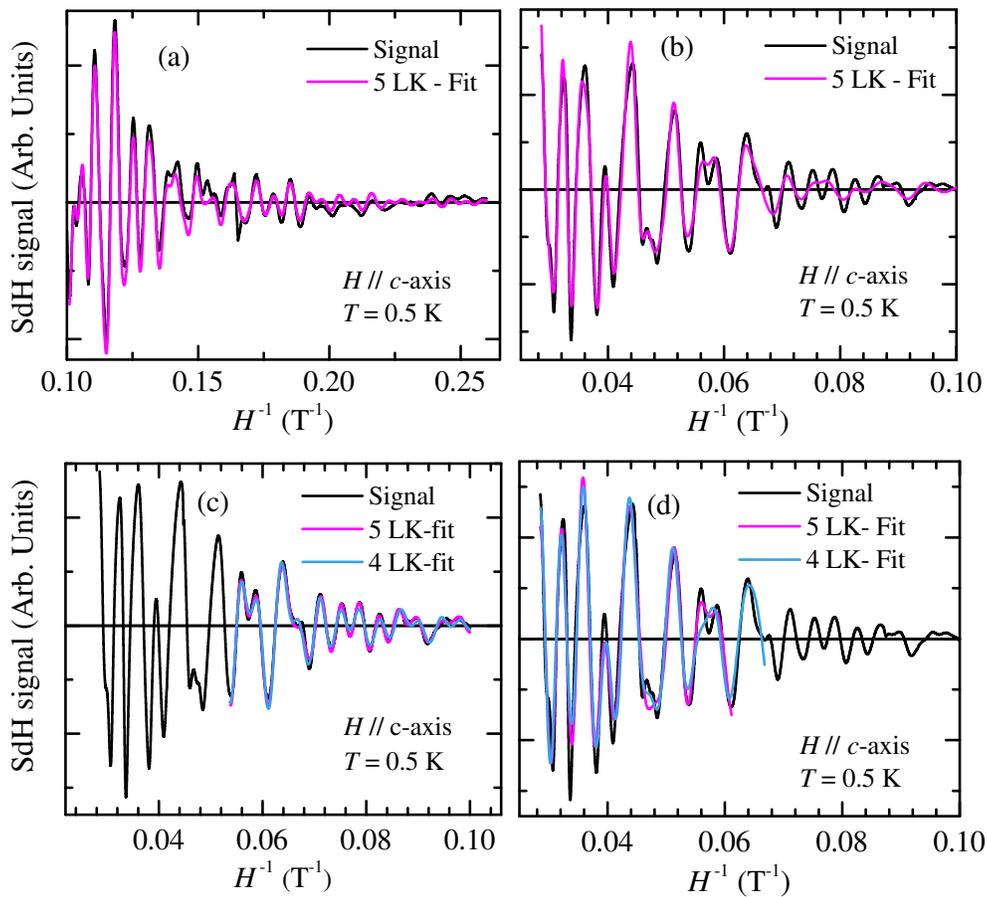}
\caption{(a) Raw SdH signal (black trace) at $T=0.5$ K as a function of the inverse field $H^{-1}$, for fields ranging between ~3.5 and 10 T applied along the \emph{c}-axis. Magenta line corresponds to a fit of the SdH signal to 5 Lifshitz-Kosevich oscillatory components, each having one of the frequencies observed in the FFT spectrum. (b) Same as in (a) but for fields ranging from 10 to 35 T. In contrast to what is observed in (a) the fit does \emph{not} correctly describe the experimental data over the entire field range. (d) Same as in (b) but over a smaller field range, i.e. from 10 to 18.6 T. Blue line depicts a fit to a 4 LK components, where one of the components corresponds to an average value between $F_{\alpha2}$ and $F_{\beta2}$. In contrast to what is seen in (b) both fits are able to reproduce the experimental data. (d) Same as in (c) but for fields ranging between 18.6 and 35 T. }
\end{center}
\label{fig:32}
\end{figure*}

A renormalization of the effective masses necessarily implies the evolution of the electronic structure at the Fermi level due to the application of an external field.
Such an evolution is likely to affect the geometry of the Fermi surface. To test this hypothesis in Fig. \ref{fig:32} we attempt to fit the oscillatory component, or the SdH signal superimposed onto the magnetoresistivity to the five frequencies observed in the FFT spectrum. However as illustrated by Figs. 5 (a) and (b) it is \emph{not} possible to fit the SdH signal over the entire field range to 5 Lifshitz-Kosevich components (4 fundamental frequencies plus $F_{\delta}$), each composed of a sinusoidal term whose amplitude is normalized by $H^{-1/2}$, the exponential Dingle factor (whose argument is effective mass and scattering rate dependent) and the temperature factor $\propto [\sinh(\alpha \mu T/H)]^{-1}$, where $\alpha$ is a constant. As seen in Fig. 5 (a), for fields below $H=10$ T one can fix the frequencies to the values obtained from the FFT spectrum and obtain a good fit of the experimental data over an extended range in $H^{-1}$ if one considers the experimental noise and a less than perfect background subtraction. However, as seen in Fig. 5 (b) if one attempts the same exercise for fields beyond 10 T one cannot converge to a fit describing the oscillatory signal over the entire $H^{-1}$ range. This is particularly surprising given that this window in inverse field $\Delta H^{-1} \simeq 0.714$ is shorter than the interval $\Delta H^{-1} > 0.16$ previously used for Fig. 5(a). As illustrated by Figs. 5 (c) and (d) it is nevertheless possible to converge to good fits if one splits this interval into two intervals, i.e. from 10 to 18.6 T and from 18.6 to 35 T. For fields beyond 10 T we were unable to simultaneously resolve $F_{\alpha2}$ and $F_{\beta2}$; the fits systematically converge to an average value between both frequencies and to another frequency around 400 T (curiously detected in the FFT spectra shown in the inset of Fig. 2(c)). This is likely attributable to the limited interval in $H^{-1}$. Hence, we have re-done the analysis after removing one LK component (blue traces) which does not compromise the fits. The important observation is that $F_{\alpha1}$ increases from 93 T (in the field interval $H \sim$ 3.5 to 10 T) to 103 T ($H \sim$ 18.6 to 35 T), $F_{\beta1}$ from 132 to 145 T and $(F_{\alpha2} + F_{\beta2})/2 = (150 + 164)/2= 157$ to 187 T. One could argue that the fits of the highest field oscillatory data are not excellent and hence question the validity of the obtained frequencies. However, here we have shown that it is \emph{not} possible to fit the entire field range with \emph{fixed} frequencies and effective masses: the Dingle and temperature factors also depend on the field interval chosen for the fits. Together with the previously shown effective mass renormalization, this indicates that the electronic structure of WTe$_2$ evolves with field due to the Zeeman splitting.

In Fig.~\ref{fig:4} we show Kohler plots of the resistivity $\rho$ normalized by its zero field value $\rho_0$ as a function of the square of the field $H^2$ normalized by $\rho_0$ and for fields along all three crystallographic orientations. Recently, it was claimed\cite{zhao} that WTe$_2$ would \emph{not} obey the Kohler rule\cite{pippard}, leading to an anomalous linear magnetoresistance for currents parallel to the  field. However, as seen in Fig.~\ref{fig:4}, our data clearly follows Kohler scaling with very mild deviations for fields along the $b-$axis, and only for the curves taken at higher temperatures. At lower fields, and for fields along the $a-$axis, indeed $\rho(H) \sim H$. Since for this current/field configuration the Lorentz force is nearly zero, which leads to no orbital magnetoresistivity, a very careful experimental effort, such as very precise field-alignment and homogeneous current distribution will be required to characterize this effect. The fact that WTe$_2$ nearly obeys Kohler rule for all three crystallographic orientations implies that its orbital magnetoresistivity is dominated by scattering by phonons and impurities.

\section{DFT calculations and the evolution of the Fermi surfaces}
\label{DFT}
\begin{figure}[htb]
\begin{center}
\includegraphics[width = 8.2 cm]{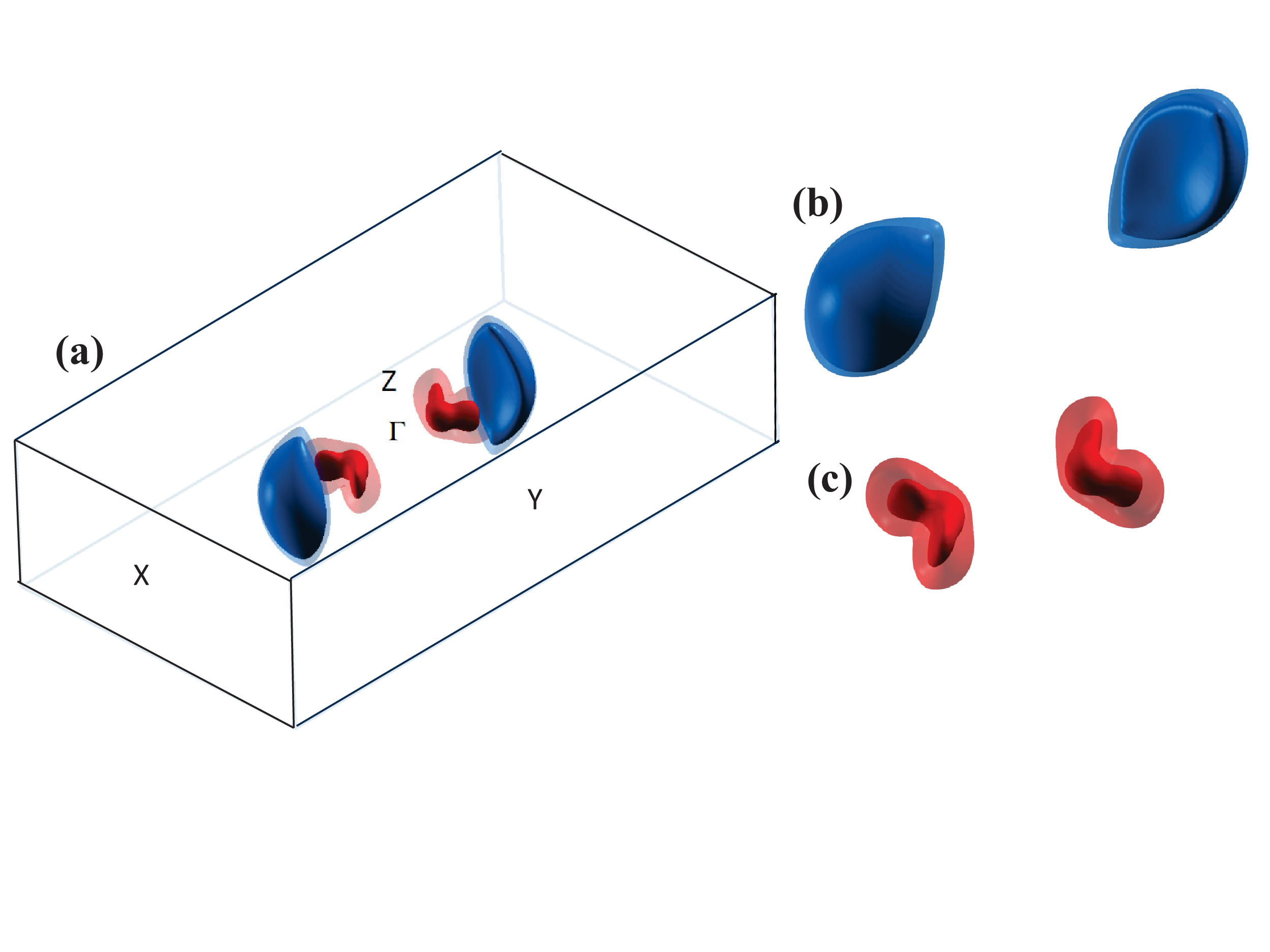}
\caption{(a) Fermi surface of WTe$_2$. The spin-orbit split electron pockets are shown in blue, and in transparent light blue color, respectively. The hole pockets are shown in red and in pink color.   The position of the energy of the electron and hole pockets was independently adjusted to obtain a good agreement with the experimentally observed Shubnikov de Haas frequencies for magnetic fields along the \emph{c}-axis.  In order to obtain high resolution, the size of the \emph{k}-point mesh used was $80 \times 80 \times 80$ . This is required to obtain, as accurately as possible, the angle dependence of the frequencies for both electron and hole orbits. The reason for 2 electron- and 2 hole-like pockets, as opposed to just two doubly degenerate ones, is the spin-orbit coupling. Therefore, the spin-orbit coupling is particularly large (relative to the energy dispersion near the Fermi level) for the hole pockets. (b) The electron pockets are shown in a higher resolution, larger image. (c) Same as in (b) but for the hole-pockets.}
\end{center}
\label{fig:5}
\end{figure}
One of the main goals of this study is to compare the geometry of the experimentally detected FSs with those resulting from \emph{ab-initio} calculations such as the ones included  in Refs.~\onlinecite{cava,pletikosic,behnia,Lv}. Notice that the initial calculations \cite{cava,pletikosic,Lv} ignored spin-orbit coupling, implying the existence of one hole- and one electron-like pocket. This would be at odds with the results reported by Ref.~\onlinecite{behnia}. In this paper, electronic structure calculations were performed by using the Vienna \emph{ab-initio} simulation package \cite{Shishkin3,Fuchs,Shishkin2,Shishkin1}(VASP) within  the generalized  gradient approximation (GGA). We have included the contribution of spin-orbit coupling in our calculations. The Perdew-Burke-Ernzerhof (PBE) exchange correlation  functional \cite{PBE} and the projected augmented wave (PAW) methodology \cite{Blochl} was used to describe the core electrons.  The 5\emph{d} and 6\emph{s} electrons for W and the 5\emph{s} and 5\emph{p} electrons for Te were treated as valence electrons in all our calculations. The energy cut off for the plane-wave basis was chosen to be 600 meV. A total of 96 bands and a $k-$point mesh of $8\times 8\times 8$ were used for the self-consistent ground state calculations. A total of 100 $k-$points were chosen between each pair of special $k-$points in the Brillouin-zone for the band-structure calculations.

In order to obtain an accurate and detailed representation of the Fermi surface we have used a $80\times  80\times 80$ k-point mesh. The areas of the observed and calculated FS orbits are related to the SdH frequencies by the Onsager relation,
\begin{equation}
F_{\vec k}  = {{\phi_0} \over {2\pi^2}} A_{\vec k}
\end{equation}
 where $\phi_0 = 2.07 \times 10^{-15}$ Tm$^2$ is the quantum of flux and $A_{\vec k}$ is the area of the electron orbit perpendicular to the field.  The effective masses of the electron- and the hole-pockets at these extremal orbits are determined by using the relation:
\begin{equation}
m_{\vec k}^*(\epsilon)= \frac{\hbar^{2}}{2\pi}
{{\partial A_{\vec k}(\epsilon)} \over {\partial  \epsilon}}.
\end{equation}
We have calculated the areas of the orbits by considering  cuts of the Fermi surface with planes intersecting the Fermi surface perpendicularly to the \emph{a}-, \emph{b}- and the \emph{c}-axis at various different angles.  In an attempt to match the experimentally observed FSs with the calculated ones, we chose to adjust our calculations to the experimental cross-sections obtained for fields along the \emph{c}-axis. For this, the energy of the electron pockets were down shifted with respect to the Fermi level by an amount of 25 meV, while the hole pockets were displaced upwards by 20 meV. The fermi surfaces  were  generated  using  the eigenvalues obtained from VASP  and  were visualized  using the  XCrysden software \cite{Kokalj}. As seen in Fig.~\ref{fig:5}, the resulting Fermi surface is composed of two sets of spin-orbit split pockets at either side of the $\Gamma$-point and along the $\Gamma$-X direction, with the electron-like pockets being depicted in blue or clear blue color, and the hole-like in red or in clear and nearly transparent red.

In Table~\ref{table:1} below, we summarize the experimental results as well as those resulting from the DFT calculations. In particular, we present a comparison between the experimental effective masses $\mu$, extracted from the fits in Fig.~\ref{fig:3}, and the band masses $\mu^b$ resulting from the DFT calculations. Here, the important observation is not only the marked contrast between the band masses and the experimental effective-masses with the band masses being systematically heavier (in particular for the $\alpha$-orbits), but also their field dependence. As seen in Table~\ref{table:1}, the effective masses are strongly field dependent, increasing by nearly a factor of 2 when they are extracted from the whole field range as compared with the masses extracted from the low field region. This is a very clear indication for the role of the magnetic field in altering the electronic structure at the Fermi level and in leading to a concomitant increase in the density of states at the Fermi level $\varrho(E_F)$. For samples that start to display quantum oscillatory behavior under fields $ H <3 $ T it is remarkable, as seen in the inset of Fig.~\ref{fig:2} (c), that one cannot resolve all four peaks in the FFT spectrum when it is taken over a broad range in $H$, such as $H=10 \rightarrow 35$ T. However, as illustrated in the Supplemental Material \cite{supplemental} through simple simulations of the experimental data, this just reflects the much lower number of oscillations within this inverse field range when compared to the number of oscillations observed in fields ranging from $< 3$ T up to 9 T.
\begin{widetext}
\begin{center}
\begin{table}[h]
\begin{tabular}[c]{c||c|c|c|c|c|c|c|c|c|c}
 & $\mu^{b}_{a}$ & $\mu_{a}$ $(7.5\rightarrow35)$ T&$\mu^{b}_{b}$ & $\mu_b$ $(3\rightarrow13)$ T& $\mu_b$ $(4\rightarrow 35)$ T & $\mu^{b}_{c}$ & $\mu_c$ $(3\rightarrow9)$ T& $\mu_{c}$ $(3\rightarrow 35)$ T &$n_{\text{ellip.}}$ & $n_{\text{DFT}}$  \\
  \hline
  \hline
  $F_{\alpha1}$ & 1.37  &$(1.17 \pm 0.06)$ & 0.84 & $(0.44 \pm 0.11)$ & $(0.97 \pm 0.04)$ & 0.72  & $( 0.42\pm 0.01 )$ & $( 0.79\pm0.04 )$ & 0.57 &  0.78 \\
  $F_{\alpha2}$ & 1.32 & $(1.20 \pm 0.07)$ & 1.02 & $(0.44 \pm 0.05)$ & $(1.13 \pm 0.05)$ & 1.1 & $( 0.47\pm 0.02)$ & $( 0.95\pm 0.05 )$ & 1.14 & 2.31 \\
  $F_{\beta1}$ & 1.02  & $(0.64 \pm 0.14)$ & 0.56  & $( 0.49 \pm 0.01 )$ & $(0.84\pm 0.03 )$ & 0.43 & $(0.33 \pm 0.01)$ & $(0.62 \pm 0.03)$ & 0.84 & 3.9 \\
  $F_{\beta2}$ & 0.97  & $(0.75 \pm 0.04)$ & 0.57  & $( 0.58\pm 0.03 )$ & $(0.92\pm 0.04)$ & 0.51 & $( 0.36\pm 0.01)$ & $( 0.64\pm0.04 )$ & 1 & 5.14 \\
  \hline
  \hline
\end{tabular}
\caption{Comparison between the experimental results and the results of the
DFT-calculations.  $\mu^{b}_{a,b,c}$ are the band masses from the DFT calculations for orbits perpendicular to fields applied along all three crystallographic axes (in units of free electron mass $m_0$), $\mu _{a,b,c}$ are the experimental effective-masses, $n_{\text{ellip.}}$ are the carrier densities as estimated from a simple ellipsoidal approximation of the Fermi surfaces, and $n_{\text{DFT}}$ are the carrier densities resulting from the DFT calculations. The $n$s are expressed in units of $10^{19}$ cm$^{-3}$. }
\label{table:1}
\end{table}
\end{center}
\end{widetext}

\begin{figure}[htb]
\begin{center}
\includegraphics[width = 8.2 cm]{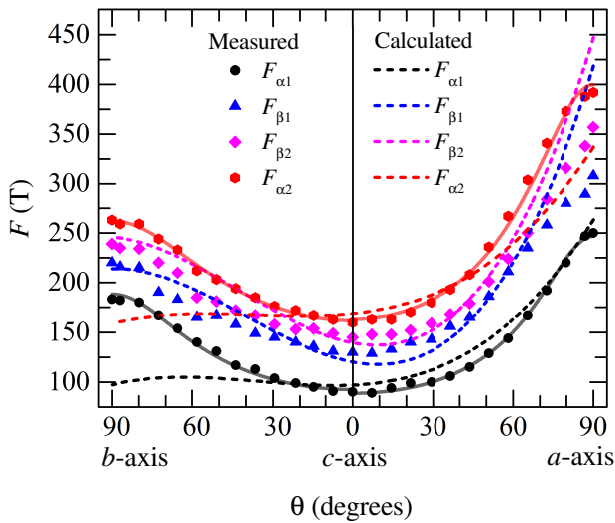}
\caption{Position of the main four peaks detected in fast Fourier spectra as a function of the orientation of the field relative to the crystallographic axes. In order to compare experimental results with the theoretical calculations dashed lines depict the DFT calculated angular dependence of the frequencies for both electron and hole pockets.  Notice the overall poor agreement with the calculations. Solid lines correspond to fits assuming ellipsoidal Fermi surfaces.}
\label{fig:6}
\end{center}
\end{figure}
Table~\ref{table:1} also shows the density of carriers $n_{\text{ellip.}}$ that one would extract from our experiments by assuming ellipsoidal FS pockets\cite{behnia}. As discussed further below, we fit the observed angular dependence of the cross-sectional areas to ellipsoidal Fermi surfaces to extract the corresponding Fermi-vectors $k_F^{a,b,c}$. The volume of each FS pocket is estimated through the expression for the volume of an ellipsoid, i.e. $V_{\text{FS}} = 4 \pi k_F^{a}k_F^{b}k_F^{c}/3$. To extract $n_{\text{ellip.}}$ one renormalizes by the volume of the Brillouin zone. This simple estimation yields a $ \sim 7$\% larger density of electrons with respect to that of the holes. However, and as we discuss below, the determination of $k_F^{a,b,c}$ is characterized by error bars in the order of $\sim 1$\%, implying an error of $\sim 3$\% in the calculation of the volume of each pocket. Hence, the frequencies or cross sectional areas extracted at low fields are consistent with near perfect compensation, if one assumed that $F_{\alpha1}$ and $F_{\alpha2}$ corresponded to hole pockets and $F_{\beta1}$ and $F_{\beta2}$ to electron pockets. Instead, if one assumed for example, that two lowest frequencies, i.e. $F_{\alpha1}$ and $F_{\beta1}$ corresponded to the hole-pockets and the other two to electron pockets, which is the indexation suggested by Ref. \onlinecite{cai}, one would obtain $\sim 50$ \% more electrons than holes thus placing this system far from compensation. The other possible combination satisfying the indexation of Ref. \onlinecite{cai} would yield $\sim 26$\% more carriers of one type with respect to the other. This would necessarily contradict the original claim in Refs. \cite{cava,pletikosic} of nearly perfect carrier compensation. Hence, our indexation is far more consistent with the pronounced magnetoresistivity displayed by this compound. The FSs calculated by DFT, and subsequently adjusted to match the experimental results, lead to a much worse situation, with the density of electrons becoming $\sim 3$ times larger than the density of holes.
\begin{figure}[htb]
\begin{center}
\includegraphics[width = 5 cm]{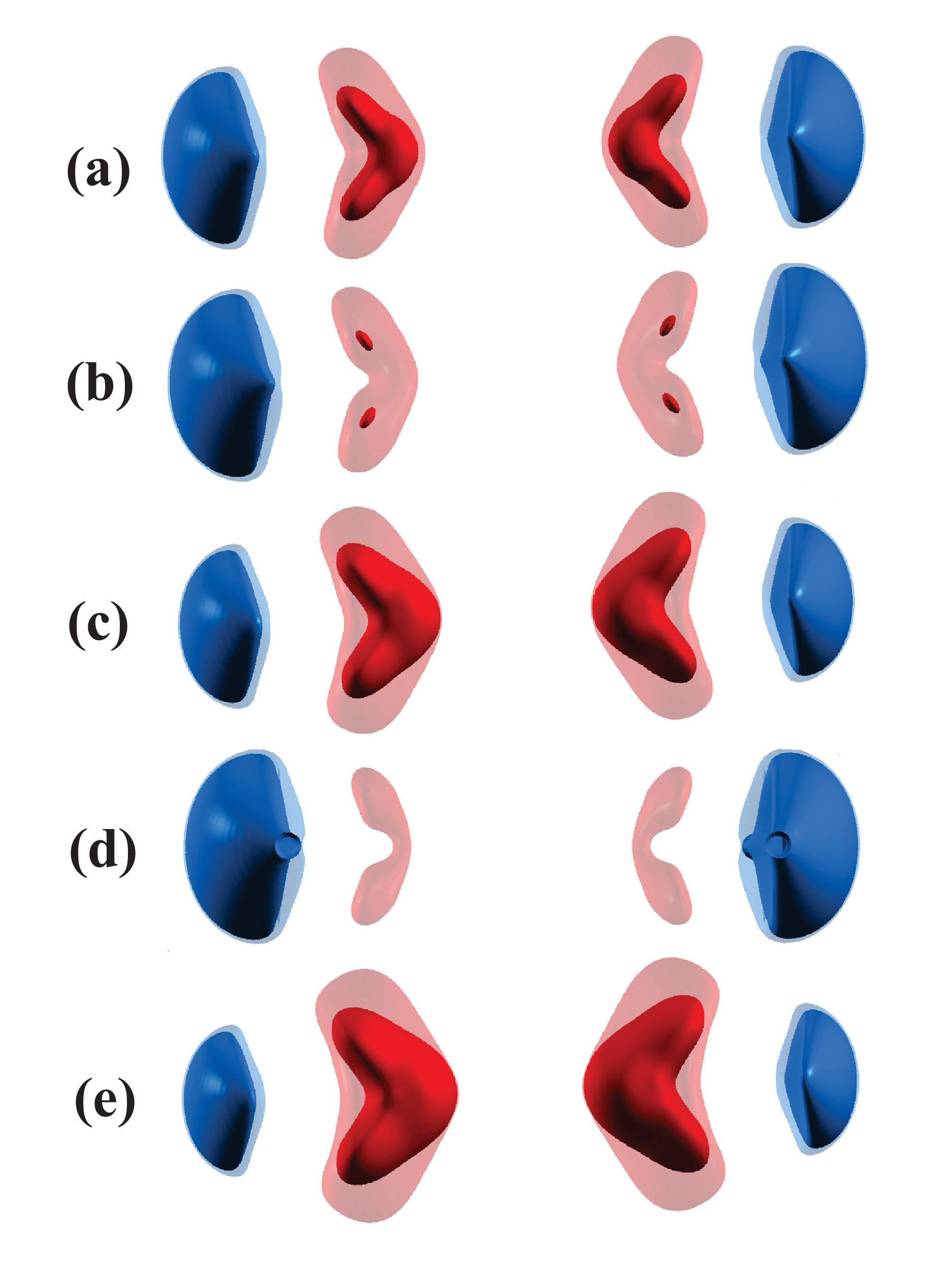}
\caption{Evolution of the Fermi surface by shifting the Fermi level by 10 and 20 meV above and below the Fermi level, respectively. (a) The electron (shown in blue and transparent light blue color) and hole pockets (shown in red and in transparent pink color) for the actual position of the Fermi level. Here, the position of the Fermi level was adjusted to obtain a good agreement with the experimental Shubnikov de Haas frequencies for magnetic fields along the \emph{c}-axis. (b) The position of the Fermi level is shifted upward by 10 meV. (c) The position of the Fermi level is shifted downward by 10 meV. (d) The position of the Fermi level is shifted upward by 20 meV. (e) The position of the Fermi level is shifted downward by 20 meV.}
\label{fig:7}
\end{center}
\end{figure}

In Fig.~\ref{fig:6}  the geometry of the FS is compared to the geometry of the DFT calculated FSs. Figure \ref{fig:6} shows the position  of the four fundamental peaks found if the FFT spectra of the SdH signal as a function of the angle relative to the crystallographic \emph{c-}axis. Here, $0^{\circ}$ corresponds to the \emph{c}-axis and $\pm90^{\circ}$ corresponds to either the \emph{a-} or the \emph{b-}axis. Dashed lines depict the angular dependence of the frequencies resulting from our DFT calculations for both the hole- and the electron-pockets including spin-orbit coupling. These curves were obtained by a polynomial fit of their calculated values at discrete angles. Our calculations yield a relatively good agreement with the experimentally obtained frequencies for fields along the \emph{c}-axis, and a modest agreement with the angular dependence of the frequencies corresponding to the electron-pockets. Nevertheless, it shows a very poor agreement with the angular dependence associated with the hole-pockets. In fact the observed angular dependence imply very similar geometry for the electron- and the hole-pockets. Solid lines correspond to fits of the observed angular dependence assuming ellipsoidal FSs. As seen, and as already pointed out by Ref.~\onlinecite{behnia}, the experimental data can be well described in terms of ellipsoidal pockets. The corresponding Fermi vectors $k_F^{a,b,c}$ extracted from our fits are given in Table II. Typical error bars in $k_F^{a,b,c}$ are in the order of 1\%. It is however quite remarkable that the DFT calculations fail to describe the electronic structure of this weakly correlated binary compound.

\begin{center}
\begin{table}[h]
\begin{tabular}[c]{c||c|c|c|c}
 $F$ & $k_{F}^{a}$ ({\AA}$^{-1})$ & $k_{F}^{b}$ ({\AA}$^{-1})$ T&$k_{F}^{c}$ ({\AA}$^{-1})$ & $V_{\text{FS}}$ ({\AA}$^{-3}$)  \\
  \hline
  \hline
  $F_{\alpha1}$ &  0.04474 & 0.06112& 0.12428& 0.0014237  \\
  $F_{\alpha2}$ &  0.05668& 0.08578& 0.13886& 0.0020823   \\
  $F_{\beta1}$ & 0.05312  & 0.07437& 0.12585& 0.0024678   \\
  $F_{\beta2}$ & 0.05431  & 0.08112& 0.13372 & 0.0028278  \\
  \hline
  \hline
\end{tabular}
\caption{Fermi vectors and Fermi surface volumes resulting from the fits of the SdH frequencies as a function of the angle assuming ellipsoidal pockets.
The error bars in the determination of $K_{F}^{a,b,c}$ are in order of 1 \%.}
\label{table:1}
\end{table}
\end{center}

In Fig.~\ref{fig:7} we illustrate the effect on the geometry of the FSs of displacing the position of the Fermi-level by 10 and 20 meV both above and below the Fermi surface.  These displacements are intended to give us a rough idea of the role played by the
Zeeman-effect on the geometry of the spin-orbit split FSs. The role of the Zeeman term is different for each one of the spin-orbit-coupling split pockets and different for electron and hole pockets and this effect is reversed when the magnetic field is applied
in the opposite direction. Furthermore, the amount of the Zeeman splitting depends on $\vec k$ since the mixing of the atomic orbitals in forming the Bloch states changes for different values of $\vec k$. Thus, we cannot infer the precise role of the Zeeman term by simply shifting the Fermi level as in Fig.~\ref{fig:7}. Nevertheless the results shown in Fig.~\ref{fig:7} conveys the sensitivity of the geometry of Fermi surfaces sheets with respect to an applied field. Notice that the hole-pockets change significantly by displacing the Fermi level by a small amount in energy, i.e., of the order of just 10  meV, which would correspond to a Zeeman splitting of the order of 100 T. The change is more dramatic when the Fermi level is shifted upwards. This exercise implies that the Zeeman-effect could induce pronounced changes in the geometry of the FSs altering the balance between opposite spin carriers. Notice also how the hole-pockets extracted from the DFT calculations, which present a boomerang like cross-section, nearly touch at its vertex when the Fermi level is displaced. Non-concentric Fermi surfaces like these would favor the magnetic breakdown between both orbits as experimentally seen.

\section{Summary}
In summary, by analyzing the angular dependence of the Shubnikov-de-Haas oscillations in single-crystals, we detect four Fermi surface cross-sectional areas in the semi-metal WTe$_2$. By following density functional theory calculations, we conclude that they correspond to spin-orbit split electron- and hole-pockets or two sets of concentric pockets at either side of the $\Gamma$-point. Nevertheless, the experimentally observed angular dependence for the cross-sectional areas is poorly described by the DFT calculations. Their geometry is well-described in terms of ellipsoidal pockets as suggested by Ref. \onlinecite{behnia}. The discrepancy between the \emph{ab-initio} calculations and the experimentally determined cross-sectional areas is far more pronounced for the hole-pockets. We observed four SdH frequencies (and not three as claimed by Ref. \onlinecite{xiang_three_pockets}) which are in excellent agreement with those reported in Refs. \onlinecite{behnia,cai}, indicating that the geometry of the extracted Fermi surfaces does \emph{not} dependent on sample quality (for samples with resistivity ratios between $10^2 < \rho(T = 300 \text{K})/ \rho(T = 0.5 \text{K}) < 2 \times 10^3$). We, as well as Ref~\onlinecite{behnia}, attribute the two middle frequencies to electron-pocket(s), with the other two corresponding to the hole one(s).

In contrast to Ref~\onlinecite{behnia}, we attribute this geometry to the spin-orbit splitting of the hole and electron Fermi-surfaces. On the other hand Ref.~\onlinecite{cai}, based on their pressure dependent studies combined with band structure calculations, attributes the middle two frequencies $F_{\beta1}$ and $F_{\beta2}$ to electron- and hole-orbits, respectively. Based on our angular-dependent study, this indexation would necessarily imply a sizeable difference in volumes between the electron- and the hole- Fermi surfaces, and therefore result in a pronounced lack of carrier compensation at zero-field. Here, we emphasize that our choice for the indexation of the Fermi surfaces in combination with the volumes of the extracted ellipsoidal Fermi surfaces, do \emph{not} discard near perfect compensation between electron- and hole-densities at zero- or low fields.

By performing measurements up to higher fields we find experimentally that the effective masses in WTe$_2$ are renormalized by the magnetic field, increasing considerably as the field increases. This is consistent with our fits of the oscillatory SdH signal as a function of the inverse field (to several Lifshitz-Kosevich components). These fits also indicate that the fundamental frequencies are evolving with field. Both observations point towards a magnetic field induced modification of the electronic structure at the Fermi level or a Fermi surface that is responsive to a magnetic field. Our DFT calculations as well as those in Ref. \onlinecite{bernevig} indicate that the geometry of the Fermi surface in WTe$_2$ is very sensitive to the exact position of the Fermi level. Therefore, it is not surprising to find that the spin-up and spin-down Fermi surfaces can be slightly modified by the Zeeman-effect, which apparently is displacing the (spin-dependent) density of states at the Fermi-level towards van-Hove singularities.
We believe that this effect could have a relevant influence on the observed colossal magnetoresistivity of WTe$_2$.

In a 3D Landau-Fermi liquid, the quasiparticle lifetime decreases with increasing quasiparticle energy
difference from the Fermi energy \cite{pines}.  If, upon application of a strong magnetic field the Fermi surface was altered, one should expect a strong influence on the quasiparticle scattering rates and lifetimes, contributing perhaps in a non-trivial manner to the large magnetoresistance observed in WTe$_2$.

Finally, the lack of agreement between our experimental observations and the band structure calculations has important implications. For example, the absence of a precise knowledge of the electronic structure at the Fermi level makes it difficult, or nearly impossible, to predict its topological properties \cite{bernevig}.

\section{Acknowledgements}
The NHMFL is supported by NSF through NSF-DMR-1157490 and the
State of Florida. LB is supported by DOE-BES through award DE-SC0002613 and by
Army Research Office through the MURI grant W911NF-11-10362.

\end{document}